\documentclass{ifacconf}
\usepackage{mathtools}
\usepackage{graphicx}      
\usepackage{natbib}        
\usepackage{algorithm}
\usepackage{algpseudocode}
\usepackage{multirow}
\usepackage{amsmath}
\usepackage{svg}
\usepackage{amsfonts}
\usepackage{pdfpages}
\usepackage{tabularx}
\usepackage{makecell}

\DeclareMathOperator*{\argmin}{arg\,min}

\begin{document}
\begin{frontmatter}

\title{Safety Filtering for Reinforcement Learning-based Adaptive Cruise Control} 


\author[First]{Habtamu Hailemichael} 
\author[First]{Beshah Ayalew} 
\author[First]{Lindsey Kerbel}
\author[Second]{Andrej Ivanco}
\author[Second]{Keith Loiselle}

\address[First]{Automotive Engineering, Clemson University, Greenville, SC 29607, USA (hhailem, beshah, lsutto2)@clemson.edu.}
\address[Second]{Allison Transmission Inc., One Allison Way, Indianapolis, IN, 46222, USA (andrej.ivanco, keith.loiselle)@allisontransmission.com}

\begin{abstract}                
Reinforcement learning (RL)-based adaptive cruise control systems (ACC) that learn and adapt to road, traffic and vehicle conditions are attractive for enhancing vehicle energy efficiency and traffic flow. However, the application of RL in safety critical systems such as ACC requires strong safety guarantees which are difficult to achieve with learning agents that have a fundamental need to explore. In this paper, we derive control barrier functions as safety filters that allow an RL-based ACC controller to explore freely within a collision safe set. Specifically, we derive control barrier functions for high relative degree nonlinear systems to take into account inertia effects relevant to commercial vehicles. We also outline an algorithm for accommodating actuation saturation with these barrier functions. While any RL algorithm can be used as the performance ACC controller together with these filters, we implement the Maximum A Posteriori Policy Optimization (MPO) algorithm with a hybrid action space that learns fuel optimal gear selection and torque control policies. The safety filtering RL approach is contrasted with a reward shaping RL approach that only learns to avoid collisions after sufficient training. Evaluations on different drive cycles demonstrate significant improvements in fuel economy with the proposed approach compared to baseline ACC algorithms.
\end{abstract}


\end{frontmatter}

\section{Introduction}
Adaptive cruise control (ACC) systems are one of the increasingly prevalent driver assistance systems for modern vehicles. An ACC system uses radar, computer vision, or laser to understand the vehicle's surrounding and make control decisions. When another vehicle or object is not in the sensing range, ACC compensates for the road grade, friction, and aerodynamic resistances to maintain a speed set by the driver. When another car or object is in front, the ACC makes decisions to prevent collision and follow the preceding vehicle as close as possible to avoid cut-ins. ACC has been shown to decrease a driver's workload, and make traffic flows efficient and safer \citep{Marsden2001,Lang2014}.

An effective ACC system should balance the traffic condition of the road, the vehicle performance, and the driver's demanded velocity. Currently available PID-based ACC systems \citep{Canale2003,Chamraz2018} and proposed MPC-based approaches \citep{Naus2008,Yang2021} are often tuned to balance this trade-off for various operating environments. Although 'adaptive' or gain-scheduled versions \citep{Radke1987} can be sought, the fixed structure of these approaches limits full adaptation throughout the lifetime of the vehicle. Furthermore, MPC-based ACC also has to find a reliable way of predicting the motion of the leading vehicle for the future horizon. On the other hand, data-driven reinforcement learning (RL) approaches offer a mechanism to continuously customize to traffic, road and vehicle conditions without a predefined control architecture \citep{Li2020}. In this work, we consider applications of RL-based ACC to commercial vehicles. In addition, while traditional ACC is primarily about the two tasks of speed tracking and maintaining a safe gap, we consider RL-based ACC (RL ACC for short) to explicitly optimize fuel economy via gear selection and torque control policies. 

Despite the potential benefits of adaptability and improved performance, RL ACC faces critical safety challenges. These derive from the needs of RL algorithms to explore in order to learn the optimal policies. RL learns how good the given state-action pair is after experiencing it, but for applications like vehicle control, exploration in an unsafe domain is unacceptable even during (on-road) training of the RL algorithms. However, thanks to recent progress in safe RL, different approaches are suggested to encourage or limit the exploration only in the safe domain. We briefly mention a few of them. Reward shaping approaches put large penalties into the performance objective function if constraints are violated. On the other hand, constrained Markov decision process (CMDP) approaches assign safety constraint costs to each state-action pair and limit the total safety constraint cost of a trajectory to be lower than a certain threshold \citep{Altman}. The reward shaping and CMDP approaches are implemented on the performance controller itself to encourage respecting safety constraints but they do not guarantee safety. Another set of approaches involve the use of safety filters that impose hard constraints. Such approaches separate the performance-oriented RL controller, whose only aim is to optimize the system's performance objective function, from the safety filters, which project the unsafe actions proposed by the performance controller into the safe set. The safety filters determine the safety condition of the given state-action pair using the dynamical model of the system, or they use offline data to learn constraints \citep{Dalal2018} and safety indexes \citep{Thananjeyan2021,Srinivasan2020}. In this paper, we pursue dynamical model-based safety guarantees to construct the safe set in such a way that gives the RL performance controller the freedom to explore within the safe boundaries. As its training progresses, the RL performance controller eventually learns the safety boundaries and ceases to demand unsafe actions \citep{Thananjeyan2021}. Note that even though it does not interfere with the inner workings, the safety filter affects control performance by dictating where the performance controller can operate.

Of the model-based approaches to designing safety filters, control barrier functions (CBFs) offer light computation and scalability \citep{li2021comparison}. A CBF guarantees safety by making the controller work in the invariant safe-set defined by a superlevel set of a continuously differentiable function $h(x):\mathbb{R}^{n} \rightarrow  \mathbb{R}$. The actions selected by the performance controllers are projected into the safe set in such a manner that the proposed actions are modified minimally \citep{Ames2019}, and no unsafe actions are passed to the controlled system. Different approaches could be pursued to specify CBFs with their pros and cons. The intuitive one is to come up with a handcrafted CBF considering the dynamics of the system and the action bounds associated with it \citep{Xu2018,Ames2014,Cheng2019}. In collision avoidance problems, for instance, the CBF can be derived by considering the maximum deceleration that the system could exert to close a distance gap. When possible, it is also desirable to progressively widen the safe set to get the maximal safe domain, a task currently possible with polynomial plant dynamics and polynomial CBFs via sum-of-squares (SOS) programming \citep{Chamraz2018}. Another approach that is tailored to high relative degree nonlinear dynamical systems such as those involving inertia effects is the use of exponential CBF (ECBF) \citep{Nguyen2016}. In this work, we derive ECBFs to work as safety filters with our RL-ACC controllers, thereby taking explicit considerations of inertia effects that are important for commercial vehicles that experience large changes in loading. 

The main contributions of this paper are then the derivation and demonstration of CBF-based safe RL-ACC approach for commercial vehicles that optimizes fuel economy. While we derive ECBFs for safety certification, we note that straight ECBFs (or CBFs in general) assume unbounded actions, and in their natural form, they might request actions that are not feasible for the vehicle's powertrain to meet. We therefore put forward a method to provide a safety guarantee for a given parameters of ECBF within the vehicle action limits. Our performance RL-ACC coordinates traction torque control and gear decisions considering fuel consumption optimization objectives. The RL ACC augmented with the safety certificate is trained and evaluated on different driving cycles, and the vehicle performance is compared with an RL ACC with reward-shaping approach to safe RL, as well as with a conventional PID-based ACC. 

The rest of the paper is organized as follows. Section \ref{sec: II} describes our derivation of the ECBF as safety filters for ACC and detail how we address actuation constraints within them. Section \ref{sec: III} describes the algorithmic details of our performance RL-ACC. Section \ref{sec: IV} discusses results and discussions, and Section \ref{sec: V} concludes the paper.

\section{Safety Filter for ACC} \label{sec: II}
We briefly review the definition of CBFs as follows. Details are given in \cite{Hsu2015}. Consider a nonlinear control affine system:
\begin{equation} \label{eq:affinedynamics}
{{\dot{x}=f\left(x\right)+g\left(x\right)u,}}.
\end{equation}
where $f$ and $g$ are locally Lipschitz, $ x\in\mathcal{R}^n $ is the system state, $ u\in\mathcal{R}^m $ is the control inputs. Assume a safe set defined by $\mathcal{C}=\left\{x\in\mathcal{R}^n|h\left(x\right)\geq0\right\}$, where  $h:\mathcal{R}^n\rightarrow\mathcal{R}$ is a continuously differentiable function. Then $h$ is a control barrier function (CBF) if there exists an extended class $ \kappa_\infty$ function $\alpha$ such that for all $x\in Int\left(\mathcal{C}\right)=\left\{x\in\mathcal{R}^n:h\left(x\right)>0\right\}:$
\begin{equation} \label{eq:CBF}
{{
\displaystyle\sup_{u\in U}{\left[L_fh\left(x\right)+L_gh\left(x\right)u\right]}\geq-\alpha\left(h\left(x\right)\right)
}}.
\end{equation}
For high relative degree nonlinear affine systems, feedback linearization could be used to develop exponential CBFs (ECBF) as detailed in \cite{Nguyen2016}. This is accomplished by transforming (input-output linearizing) the high relative degree nonlinear systems into a virtual linear system with new state variable $\eta_b := [h(x),\dot{h}(x), \cdot \: \cdot \: \cdot,h^r (x)]^T$, input $ \mu$ and output $ h\left(x\right)$:
\begin{equation} \label{eq:statefeedbacklinearizaion}
 \begin{aligned}
 \dot\eta_{b} &= F\eta_{b}\left(x\right)+G\mu, \\
 h\left(x\right) &= C\eta_{b}
\end{aligned}
\end{equation}
where $F$ and $G$ are matrices representing an integrator chain, and $C=[1,\: 0, \: \cdot \: \cdot\: \cdot \:, 0]$. A state feedback controller can be designed for the transformed system as: $\mu={-K}_\alpha\eta_b$ with a suitable gain vector $K_\alpha$ that makes $ F - GK_{\alpha}$ Hurwitz. For a system with relative degree $r$, $\mu$ is also $r^{th}$ derivative of the output $h(x)$, $\mu =L_f^{r}h(x) + L_g\L_f^{r-1}h(x)u$. If there exists a state feedback gain $K_\alpha$ that makes $\mu\geq-K_\alpha\eta_b\left(x\right)$ for all states, then one can show that $h(x)$ is an exponential control barrier function (see \cite{Nguyen2016}). 

The ACC part of the present problem is modelled with the state variables of separation distance $z$, the velocity of the host vehicle $v_h$ and the velocity of the leading vehicle $v_l$. The corresponding state equations are: 
\begin{subequations}\label{eq:ACCsetup}
  \begin{equation}
    \label{eq:ACCsetup-a}
      {{\dot{z}=v_l-v_h}}
  \end{equation}
  \begin{equation}
    \label{eq:ACCsetup-b}
    {{\dot{v_l}=a_l}}
  \end{equation}
  \begin{equation}
    \label{eq:ACCsetup-c}
    {{{\dot{v}}_h=\frac{T_t}{r_wm_v}-\frac{F_r\left(v_h,m_v,\theta\right)}{m_v}}}
  \end{equation}
\end{subequations}

\begin{equation} \label{eq:resistanceforces}
{{
F_r=\frac{\rho A c_dv_h^2}{2}+m_vgf\cos{\theta}+m_vg\sin{\theta}
}}
\end{equation}

where $F_r$ is the total resistance force including gravitational, rolling and aerodynamic resistances, and $T_t$ is the traction torque at the wheels. The parameters $c_d$, $f$, $\theta$, $m_v$ , $\rho$, $A_v$, $r_w$, $a_l$ are aerodynamic coefficient, rolling resistance coefficient, road grade, mass of the vehicle, density of air, frontal area of the vehicle, radius of the wheels, and acceleration of the leading vehicle, respectively.

We observe that the above model can be readily put in the control affine form (\ref{eq:affinedynamics}). Given a collision safety objective, we seek the separation distance $z$ to always be above a specified minimum inter-vehicle distance $z_0$. To this end, we define the control barrier function (CBF) as the output $h\left(x\right) = z-z_{0}$. Considering that the control actuation is the traction torque $T_t$, we have a control affine system of relative degree two. In physical terms, the safety objective is on position while traction torque directly manipulates acceleration. Inertia effects come into play and must be accounted for. The input-output linearization into the form (\ref{eq:statefeedbacklinearizaion}) then gives:

\begin{equation} \label{eq:hx}
{{
\dot{h}(x) = v_l-v_h,
}}
\end{equation}
\begin{equation} \label{eq:hddot}
{{
\mu=\ddot{h}\left(x\right)=\frac{F_r\left(v_h,m_v,\theta\right)}{m_v}+a_l-\frac{T_t}{m_vr_w},
}}
\end{equation}
\begin{equation} \label{eq:kaeta}
{{
{-K}_\alpha\eta_b\left(x\right) = -k_{\alpha1}\left(z-z_0\right)\ -k_{\alpha2}\left(v_l{-v}_h\right)
}}
\end{equation}

We now compute some bounds for the given control input $\mu$ considering actuation limits on the traction torque ($T_{min}$ and $T_{max}$). For a given acceleration of the preceding vehicle $\left(a_l\right)$ and velocity of the host ($v_h$), the feasible bounds of $\mu$ are given as
\begin{equation} \label{eq:muminmax}
{{
\mu_{T_{min/max}}=a_l+\frac{F_r\left(v_h,\theta,m_v\right)}{m_v}-\frac{T_{min/max}}{m_vr_w}
}}
\end{equation}
For a given gain vector ${K_\alpha=[k}_{\alpha1},\ k_{\alpha2}]$, ECBF guarantees safety if the proposed state feedback control, $ {-k}_{\alpha1}\left(z-z_0\right)-k_{\alpha2}\left(v_l-v_h\right)$, is within the virtual linear system action bound $\left[\mu_{Tmax},\ \mu_{Tmin}\right]$. In general application cases, however, this bound may not be respected. Nevertheless, if $K_\alpha$ is chosen so that the poles are placed sufficiently to the left in s-plane, the above ECBF could still bound the safe set. Safety assurance for such pole selections could be achieved by investigating the evolution of the CBF control term $h\left(x\right)$ in worst-case situation where the linear virtual model is initialized with extreme possible $\eta_{0,xrm}$, and then the possible limiting torque actions are applied. For a given minimum separation distance target and maximum downhill road grade, this is equivalent to applying the maximum possible traction torque output of the performance RL-ACC agent, with the host vehicle model (of largest loading) initialized in with the maximum possible velocity while the preceding vehicle is under its maximum deceleration. This extreme conditions gives the feasible $\mu$ bounds as $\mu_{Tmin-xrm}$ and $\mu_{Tmax-xrm}$ using equations (\ref{eq:muminmax}).

To capture the evolution of $h\left(x\right)$ under these extreme conditions, a simulation rollout is discretized into timestep $\Delta{t}$, and the action $\mu$ (saturated with $\mu_{Tmin-xrm}$ and $\mu_{Tmax-xrm}$) held piecewise constant. Algorithm \ref{alg:cap} shows how this is implemented by integrating the virtual system (3). If the $h\left(x\right)$ from this simulation is positive at infinity (or after some finite time), the selected $K_\alpha$ guarantees safety. Otherwise, the $K_{\alpha}$ needs to be changed until this is satisfied.

\begin{algorithm}
\caption{An algorithm to enforce system bounds on a virtual linear system}\label{alg:cap}
\begin{algorithmic}
\State $\eta \gets \eta_0$
\State $\mu \gets \mu_0$
\While{$t \leq t_{\infty}$}
\State $t \gets t + \Delta t$
\If{$\ \mu\ <\ \mu_{Tmax-xrm}$}
    \State $\mu \gets \mu_{Tmax-xrm}$
\ElsIf{$ \mu\ > \mu_{T{min-xrm}}$}
    \State $\mu \leftarrow  \mu_{T{min-xrm}}$
\EndIf
 \ \State $\ h\left(x(t)\right)\gets C(e^{F\Delta{t}}\eta_0+ e^{F\Delta{t}}\int_{0}^{\Delta t}e^{-F \tau}G\mu d(\tau))$
 \ \State $ \ \mu\ \gets -k_{\alpha1}h\left(x\right)-\ k_{\alpha2}\dot{h}\left(x\right)$
  \ \State $\  \eta_0 \gets \left[\begin{matrix}h\left(x\right)\\ \dot{h}\left(x\right)\\\end{matrix}\right]$
\EndWhile
\end{algorithmic}
\end{algorithm}

Once the suitable gain vector $K_\alpha$ is selected, the ECBF safety constraint enforces safety by projecting the action proposed by the outputs of the RL controller’s actor network $T_a\left(s\right)$ (see next section) to the control traction torque $T_t$ in a way that introduces minimal changes to it. This is done by posing and solving the quadratic program:
\begin{equation}
\label{eq:optimization}
\begin{aligned}
 T^{*}_t =  & \displaystyle\argmin_{T_{t}} \frac{1}{2}\left \| T_{t}-T_{a}(s) \right \|^2 \\
\textrm{s.t.} \quad & a_l+\frac{F_r\left(v_h,m_v,\theta\right)}{m_v}-\frac{T_t}{m_vr_w}\geq{-k}_{\alpha1}\left(z-z_0\right)\\
\quad &  - k_{\alpha2}\left(v_l-v_h\right)\\
\end{aligned}
\end{equation}

\section{Vehicle Environment and RL ACC } \label{sec: III}
\begin{figure}
\begin{center}
\includegraphics[width=8.4cm, height=5.5 cm]{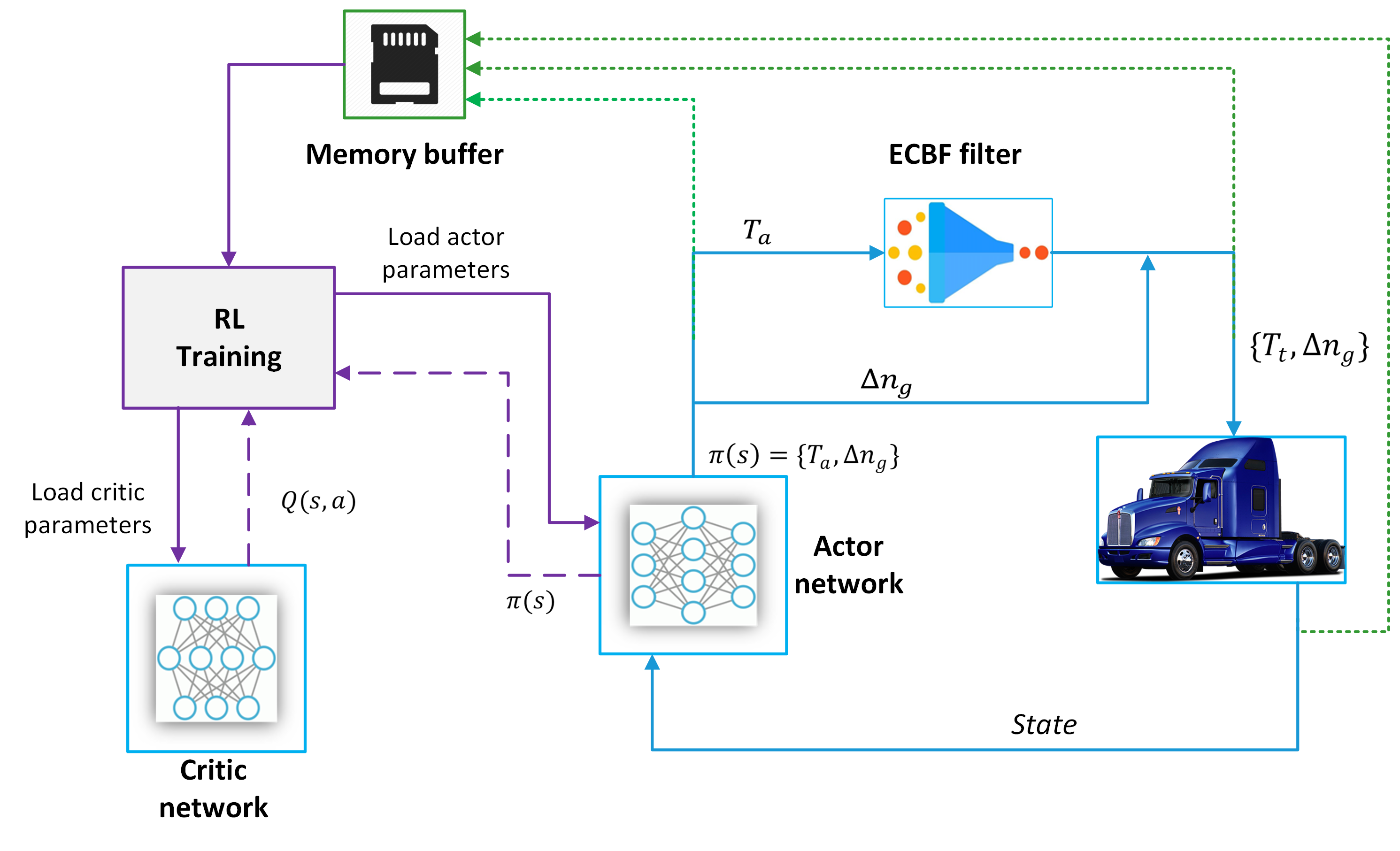}    
\caption{Training RL agent for ACC} 
\label{fig:Training}
\end{center}
\end{figure}

The powertrain controller is modeled as Markov decision process (MDP) consisting of states $s$, actions $a$, a reward function $r\left(s,a\right)$, and discounting factor $\gamma$. The probability of action choices is policy $\pi(a|s,{\boldsymbol\theta})$ where ${\boldsymbol\theta}$ denotes the parameters of the deep neural network used to approximate the policy. The host vehicle velocity $v_l$, the relative velocity between the preceding and host vehicles $v_{rel}$, the separation distance between the vehicles $z$, the gear $n_g$, the mass of the vehicle $m_v$, the road grade $\theta$, the driver demanded velocity $v_{set}$ and a flag to show if the vehicle is in ACC sensor range$\ f$ constitute the states of the RL agent, $s=\{v_l,v_{rel},z,n_{g,}m_v,\theta,v_{set},f\}$. The RL performance controller is designed to perform both traction torque $T_a$ control and gear change selection $\Delta{n_g}$, i.e. $a=\{T_a,\Delta{n_g}\}$. As shown in Fig.\ref{fig:Training}, the proposed $T_a$ is filtered by the ECBF safety layer to safe traction torque demand $T_t$ (\ref{eq:optimization}). The engine torque and engine speed that brings about this wheel traction torque are then calculated utilizing transmission ratios of the selected gear and the final drive, and the associated fuel rate is read from the fuel map. Notice that while the RL controller's actions are $T_{a}$ and $\Delta{n_g}$, the ECBF safety filter does not use $\Delta{n_g}$ in the safety constraint. However, taking into account that gear selection is crucial for fuel economy and driver accommodation, it is an integral part of the RL performance controller.

The filtered traction torque $T_t$ and the gear change $\Delta n_g$ actions are implemented in the vehicle environment, and the suitability of the actions is measured by the reward function. The reward is designed to accomplish the in range and out of range tasks, and different performance objectives within each task are tuned by reward weights ($w$). When there is not a vehicle present in the sensing range $\left(z>z_{sr}\right)$, as shown in (\ref{eq:outofrange}), the reward structure requires the vehicle to maintain the driver-set velocity and concurrently balances the fuel consumption and smooth torque change considerations. When there is a vehicle in the sensing range, on the other hand, the reward aims to maintain a close distance from the preceding vehicle, as shown in (\ref{eq:inrange}). In such proximity, in addition to smooth torque change and fuel consumption considerations, the reward $r_{os}$ discourages the host vehicle from overspeeding beyond the driver demanded velocity ($v_{set}$). Gear hunting and the associated rough vehicle operation are mitigated by including a gear reward term weighted by $w_g$.
\begin{equation} \label{eq:outofrange}
\begin{split}
r=w_v0.1^\frac{{|v}_h-v_{set}|}{V_{rel,max}}+w_f0.1^\frac{{\dot{m}}_f}{m_{f,max}} + w_T0.1^\frac{{|\Delta T_e}|}{T_{e,max}} +\\ w_{g}0.1{^\frac{{|\Delta n_{g}}|}{n_{g,max}}}
\end{split}
\end{equation}

\begin{equation} \label{eq:inrange}
\begin{split}
r=\ w_{z}0.1^{\frac{Z}{Z_{sr}}}+w_{f}0.1^{\frac{{\dot{m}}_f}{m_{f,max}}}+w_{T}0.1^{\frac{{|\Delta T_e}|}{T_{e,max}}}+\\
w_{g}0.1{^\frac{{|\Delta n_{g}}|}{n_{g,max}}}+r_{os}
\end{split}
\end{equation}

where
$r_{os}=w_{os}\ if\ v_h\le v_{set},\ else:{r_{os}=\ w}_{os}0.1^\frac{v_h-v_{set}}{v_{rel,max}}$,\\
 $\dot{m}_f$ is the fuel rate and $T_e$ is the engine torque.

 To accommodate the continuous traction torque and the discrete gear selection, Hybrid Maximum A Posteriori Policy Optimization (HMPO) is found to be a good fit for the RL training algorithm \citep{Kerbel2022,Neunert2020,Abdolmaleki2018}. In addition to being scalable and robust like state of the art Proximal Policy Optimization (PPO) \citep{Schulman2017} and Trust-Region Policy Optimization (TRPO) \citep{Schulman2015} algorithms, the fact that it is off-policy makes it data efficient to apply it to the real world RL ACC trainings. The RL agent comprises of an actor (parameterized by ${\boldsymbol\theta}$) and a critic (parameterized by ${\boldsymbol\phi}$) networks, in which the former determines the control policy for a given state $\pi\left(s\middle|{\boldsymbol\theta}\right)$ and the latter evaluates these actions by providing the associated action values $Q\left(s,a\middle|{\boldsymbol\phi}\right)$. The actor network outputs the mean and variance of a Gaussian distribution, from which traction torque is sampled (\ref{eq:torqeguassian}). In addition to that, it uses softmax activation at the output layer with three choices for the gear change decision, analogous to the available gear changes $\Delta n=\{1,0,-1\} (upshift, no change, downshift)$. Categorical sampling is then used to obtain the gear change policy (\ref{eq:gearcatagory}). Assuming independence between the continuous $\pi_{\boldsymbol\theta}^T\left(T_a|s\right)$ and discrete $\pi_{\boldsymbol\theta}^g(\Delta{n_g}|s)$ policies, the total policy could be factorized as (\ref{eq:geartorque}) for combine action $a = \{T_a,\Delta n_g\}$.	

\begin{equation} \label{eq:torqeguassian}
{{
\pi^{T}_{\boldsymbol\theta\left(T_a|s\right)}=\mathcal{N}\left(\mu_{\boldsymbol\theta}\left(s\right),\sigma_{\boldsymbol\theta}^2\left(s\right)\right) 
}}
\end{equation}
\begin{equation} \label{eq:gearcatagory}
{{
\pi_{\boldsymbol\theta}^g(\Delta n_g|s)=Cat(\alpha_{\boldsymbol\theta}(s)),\forall s\ \sum_{k=1}^{3}{\alpha_{k,{\boldsymbol\theta}}\left(s\right)=1}
}}
\end{equation}
\begin{equation} \label{eq:geartorque}
{{
\pi_{\boldsymbol\theta}\left(a\middle| s\right)=\pi_{\boldsymbol\theta}^T\left(T_a\middle| s\right)\pi_{\boldsymbol\theta}^g(\Delta n_g|s))
}}
\end{equation}
In the policy improvement phase, MPO samples from the Q-function for different actions and update the actor-network parameters to output actions that maximize the action values $Q(s,a)$. This is accomplished by optimizing the likelihood function of acting optimally using the expectation-maximization algorithm ( see \cite{Neunert2020,Abdolmaleki2018}). The policy evaluation phase of the training fits the Q-function $Q_\theta\left(s,a,{\phi}\right)$ of the critic network, with parameters ${\boldsymbol\phi}$, by minimizing the square loss of the current $Q_\theta\left(s,a,{\boldsymbol\phi}\right)$ and a target defined by retrace sampling $Q_t^{ret}$ \citep{Munos2016}.
\begin{equation} \label{eq:Criticlearning}
{{
\displaystyle\min_{\boldsymbol\phi}{L\left(\boldsymbol\phi\right)=\displaystyle\min_{\boldsymbol\phi}{\mathbb{E}_{\left(s,a\right)\sim\mathcal{R}}\left[Q_{\boldsymbol\theta}\left(s,a|{\boldsymbol\phi}\right)-Q_t^{ret}\right]^2}}
}}
\end{equation}

\section{Results and Discussions } \label{sec: IV}
 The above RL ACC with the ECBF safety filter is applied to a model of medium duty truck in urban and highway driving conditions. The actor and critic networks are constructed with three hidden layers, and each layer consists of 256 nodes. The simulation uses a 10-speed automated manual transmission (AMT) truck that has a 5 to 10 tons weight range. The preceding vehicle follows Federal Test Procedure (FTP-75) drive cycle for the urban driving training, while for highway driving, a combination of Highway Fuel Economy (HWFET) and ArtMw130 cycles are used in succession \citep{Barlow}. Once trained, we will use different drive cycles for evaluation as will be described below.

 In each simulation step, as shown in Fig.\ref{fig:Training}, the actor network proposes the torque and the gear actions for a given state which will be filtered by the ECBF safety layer. The vehicle environment then executes the safe actions, and the associated rewards are calculated. To accommodate the different objectives of each task, the reward is structured with weights of $[w_v=0.675,w_f=0.175,w_T=0.075,w_g=0.075]$ for in range, and $[w_z=0.325,w_f=0.175,w_{os}=0.35,w_T=0.075,w_g=0.075]$ for out of range conditions. The state, action and rewards are stored in the memory buffer, and afterward, batches of these data are used to train the networks using the HMPO algorithm. In order to prevent RL from learning the specific drive cycles, the vehicles are initialized in random separation distance along with the addition of noise to the velocity profile of the preceding vehicle. The weight fluctuations are considered by varying the truck weight within and between training episodes. 

During training, because of the careful choice of the gain vector $K_\alpha=\left[0.2,5\right]$ as per section \ref{sec: II}, the vehicle never crashes nor comes within safe distance $z_0$. As the training progresses, the RL learns to operate near the driver set velocity when it is out of range and follows the preceding vehicle more and more closely when it is in range. Even though it is not provided with the engine efficiency map, as exhibited by the improvement of MPG with training, the RL network eventually learns the fuel optimal gear and torque actions. 

\begin{table}[hb]
\begin{center}
\caption{Vehicle environment and RL hyperparameter setting}\label{tb:parametersMPO}
\begin{tabular}{cc|cc}
\hline
\multicolumn{2}{| c |}{Vehicle Parameters} & \multicolumn{2}{| c |}{MPO Hyperparameters}   \\\hline
Mass & 5\ - 10 tons & Actor, critic learning rate & ${10}^{-4},{10}^{-5}$  \\
$A_u$ & $7.71m^2$ & Dual constraint&  0.1 \\
$C_d$& $0.08$ & Retrace steps& 15 \\
$r_w$& $0.498$ & KL constraints $\epsilon_\mu,\epsilon_\sigma,\epsilon_d$ & $0.1,0.001,0.1$ \\
$f$& $0.015$ & $\alpha_d,\alpha_c$ & 10\\
$z_{sr}$&$350$ & $\gamma$ & 0.99 \\\hline
\end{tabular}
\end{center}
\end{table}
Even if it is not practical for safety critical systems, a reward shaping approach of safeguarding safety is considered to compare against the ECBF-based safety filtering. A penalty of $r_s=-1$ is added to the reward function when the host approaches closer than the minimum safe distance limit $z_0$ and, in the situation of a crash, the penalty is enlarged to $r_c=-10$. Due to these safety violation penalties, unsafe actions reduce with training, and eventually, the agent learns to maximize the reward safely. In addition to the reward shaping approach, the conventional PID ACC is used as a baseline which, like in the case of RL, is designed by dividing the control into phases for the in range and out of range conditions \citep{Canale2003}. The traction torque $T_t$ request is given by PID controller and an optimal gear is chosen based on the gear with the lowest fuel rate given the desired traction torque and vehicle velocity \citep{Yoon2020,Kerbel2022}. 

After the RL ACC with ECBF is trained, its performance is evaluated and compared with PID ACC and RL ACC with reward shaping counterparts on a 9-ton truck in urban and highway driving conditions. For the urban case, the preceding vehicle follows the ArtUrban drive cycle, and the driver demanded velocity $v_{set}$ is set to be $15\ m/s$. Similarly, a $v_{set}$ of $25\ m/s$ is used for highway driving, and to better capture different velocity profiles in the highway situation, the preceding vehicle follows a combination of ArtRoad and ARTMw150. The initial separation distance between the vehicles is $1500\ m$ in both cases.

\begin{figure}[htbp]
\begin{center}
\includegraphics[width=9.4cm,height=6 cm]{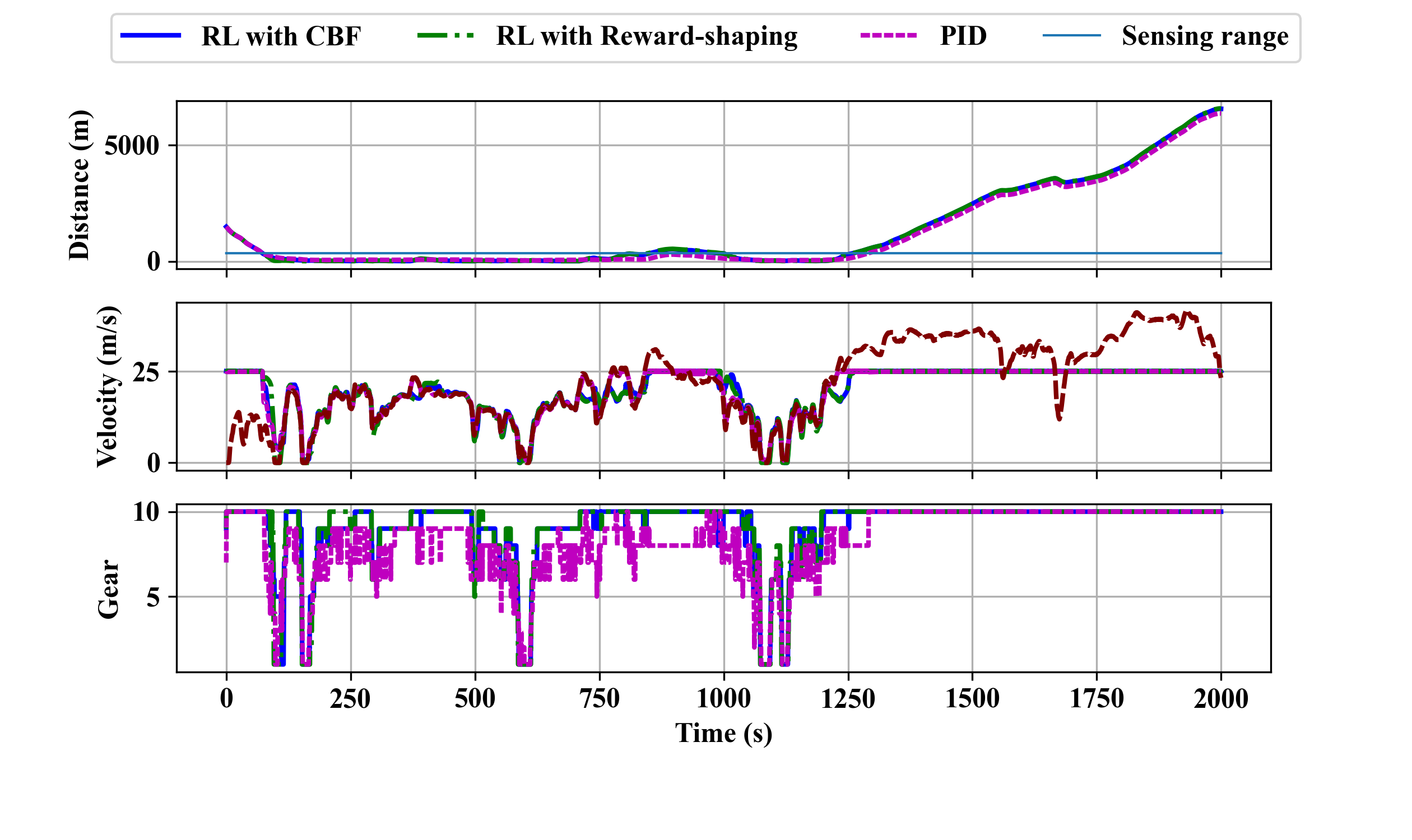} 
\vspace*{-10mm}
\caption{Simulation of separation distance, velocity, and gear profiles of RL and PID ACC controllers in a highway driving.}  
\label{fig:Highway}
\end{center}
\end{figure}
In both driving conditions, the RL ACC successfully meets the in range as well as out of range objectives and, most importantly, safety constraints are respected. Fig.\ref{fig:Highway} shows the RL ACC has a similar velocity profile to its PID ACC counterpart for the most part of the simulation. However, when it comes to gear selection, the RL ACC tends to operate at higher gears. As summarised in Table \ref{tb:Performance}, for highway driving, the RL ACC exhibited an MPG improvement of $8.3\%$, whereas, in the case of urban driving, it has $7.9\%$ higher MPG than the PID ACC baseline. When the preceding vehicle is in range, the RL ACC is less susceptible to cut-in as it follows the preceding vehicle closer, shown by the lower mean in range separation distance $z_{ir}$. Moreover, it is possible to see that the RL ACC with ECBF filter and the RL ACC with reward shaping arrangements achieve equivalent levels of fuel economy and in range car following performances.

Table \ref{tb:massflact} shows the performance comparison with weight fluctuation in which the vehicle's weight ranges from 5 to 10-tons. The RL ACC maintains higher MPG than the PID ACC throughout the given weight range, and the separation distance is not significantly influenced.

\begin{table}[hb]
\begin{center}
\caption{Performance comparison between PID ACC, RL ACC with ECBF and RL ACC with reward shaping}\label{tb:Performance}
\begin{tabular}{p{0.8cm}p{0.8cm}p{0.8cm}p{0.8cm}p{0.8cm}p{0.8cm}p{0.8cm}}
\hline
&\multicolumn{3}{| c |}{\thead{Highway driving}} & \multicolumn{3}{| c |}{\thead{Urban driving}} \\\hline
ACC  & PID & RL & RL  & PID  & RL & RL\\\hline
Safety layer & -& ECBF & Reward\par shaping & - & ECBF & Reward \par shaping\\\hline
MPG & 8.6\par (-) & 9.3\par (8.31\%) & 9.31\par (8.37\%) & 6.8\par (-) & 7.35 \par (7.9\%) & 7.38 \par (8.4\%)\\
$Z_{ir}(m)$ & $95$ & $74$ & $73$ & $42$ & $39$ & $38$\\ \hline
\end{tabular}
\end{center}
\end{table}

\begin{table}
\caption{Perandomizedof PID ACC and RL ACC with vehicle mass fluctuation}
\label{tb:massflact}

\begin{tabular}{{p{0.7cm}|p{0.7cm}|p{0.675cm}p{0.675cm}p{0.675cm}p{0.675cm}p{0.675cm}p{0.675cm}}}
\hline
 & Weight \par (tons)& 5 & 6 & 7 & 8 & 9 & 10 \\ \hline 
\multirow{2}{=}{RL with ECBF} & {MPG} & 10.58\par(10.9\%)& {10.38\par(11.6\%)}& {9.99\par(9.6\%)}& {9.61\par(8.3\%)}& {9.3\par(8.31\%)}& {8.95\par(7.6\%)}\\ 
 & {$Z_{ir}(m)$} & {67}& {69}& {73}& {75}& {74}& {77}\\\hline
 \multirow{2}{*}{PID} & {MPG} & {9.54} & {9.3}& {9.11}& {8.87}& {8.6}& {8.32}\\
 & {$Z_{ir}(m)$} & {95}& {95}& {94}& {95}& {95}& {96}\\\hline
 
\end{tabular}
\end{table}

\section{Conclusion} \label{sec: V}
In this paper, an exponential control barrier function-based safety filter is employed to instill safety into RL based ACC system by projecting the learning exploration to a safe set. Since practical systems operate with bounded actions, we proposed an approach to verify the safety of a given ECBF design by forward simulating in consideration of worst case scenarios. After being filtered by this ECBF, the traction torque and gear change actions proposed by the RL-based ACC are implemented on a simulated vehicle environment and the associated rewards are observed. The RL networks are trained using Hybrid Maximum A Posteriori Policy Optimization (HMPO) algorithm that accommodates the continuous traction torque and discrete gear change actions. Evaluation on a medium-duty truck shows that the RL ACC fulfilled the velocity objectives and, most importantly, respected the safety constraints. Compared to PID ACC, the RL ACC augments MPG by $8.3 \%$ in highway driving conditions when the preceding vehicle follows a combination of ArtRoad and ARTMw150 drive cycles, and by $7.9 \%$ in urban driving conditions when the preceding vehicle follows ArtUrban drive cycle. Moreover, the RL ACC learns to handle weight fluctuations and maintains high performance throughout the vehicle's weight range.

The current algorithm training and evaluations are performed on standard driving cycles. Future work will focus on using randomized traffic data and measurement noise to assess the performance and robustness of RL ACC in even more realistic driving conditions. In addition, future work will also look at less conservative methods of accounting for uncertainties (not worst-case) in ECBF design.




\bibliography{main}             
                                                   







\end{document}